\documentclass{sig-alternate-05-2015}
\pdfoutput=1
\usepackage{epstopdf}
\usepackage{multirow}
\usepackage{balance}
\usepackage{amsfonts}
\usepackage{amsmath}
\usepackage{amssymb}
\usepackage{graphicx}
\usepackage{subfigure}
\usepackage[noend,ruled]{algorithm2e}
\usepackage{microtype}
\usepackage{lmodern}
\usepackage{url}
\usepackage{bbm}
\usepackage{enumitem}
\usepackage[table]{xcolor}

\newtheorem{thm:def}{Definition}
\newtheorem{thm:eg}{Example}
\newtheorem{thm:lem}{Lemma}

\newcommand{\nop}[1]{}

\newcommand{\mquote}[1]{{``\emph{#1}''}}
\newcommand{\HINSim}{\mbox{\sf ESim}\xspace}
\newcommand{\PathSim}{\mbox{\sf PathSim}\xspace}
\newcommand{\LINE}{\mbox{\sf LINE}\xspace}
\newcommand{\PTE}{\mbox{\sf PTE}\xspace}

\begin{document}

\title{Meta-Path Guided Embedding for Similarity Search in Large-Scale Heterogeneous Information Networks}







\numberofauthors{1}
\author{
	Jingbo Shang$^1$, Meng Qu$^1$, Jialu Liu$^2$, Lance M. Kaplan$^3$, Jiawei Han$^1$, Jian Peng$^1$ \\
	\begin{tabular}{c}
		\affaddr{{$^1$}{Department of Computer Science, University of Illinois Urbana-Champaign, IL, USA}} \\
		\affaddr{{$^2$}{Google Research, New York City, NY, USA}} \\
		\affaddr{{$^3$}{Sensors \& Electron Devices Directorate, Army Research Laboratory, Adelphi, MD, USA}} \\
	\end{tabular} \\
	\email{\normalsize {$^1$}{\{shang7, mengqu2, hanj, jianpeng\}@illinois.edu}}
	\email{\normalsize {$^2$}{jialu@google.com}}
	\email{\normalsize {$^3$}{lance.m.kaplan.civ@mail.mil}}
}

\clubpenalty=10000
\widowpenalty = 10000
\predisplaypenalty=10000

\maketitle
\begin{abstract}

Most real-world data can be modeled as heterogeneous information networks (HINs) consisting of vertices of multiple types and their relationships.
Search for similar vertices of the same type in large HINs, such as bibliographic networks and business-review networks, is a fundamental problem with broad applications.
Although similarity search in HINs has been studied previously, most existing approaches neither explore rich semantic information embedded in the network structures nor take user's preference as a guidance.

In this paper, we re-examine similarity search in HINs and propose a novel embedding-based framework.
It models vertices as low-dimensional vectors to explore network structure-embedded similarity. 
To accommodate user preferences at defining similarity semantics, our proposed framework, \HINSim, accepts user-defined meta-paths as guidance to learn vertex vectors in a user-preferred embedding space.
Moreover, an efficient and parallel sampling-based optimization algorithm has been developed to learn embeddings in large-scale HINs.
Extensive experiments on real-world large-scale HINs demonstrate a significant improvement on the effectiveness of \HINSim over several state-of-the-art algorithms as well as its scalability.

\nop{ masked by JH @ 7/27/16
A lot of real-world data can be modeled as heterogeneous information networks (HINs) consisting of vertices of multiple types and their relationships.
Search for similar vertices of the same type in large HINs, such as bibliographic networks and business-review networks, is a fundamental problem with broad applications.
Unfortunately, most existing approaches on HIN-based similarity search neither capture sufficiently user-preferred semantic similarity, nor take good consideration of information propagation over a network.

In this paper, we re-examine this problem and propose a novel framework for similarity search in large-scale HINs.
It models vertices as low-dimensional vectors to support highly efficient similarity queries.
To accommodate user preferences and resolve the ambiguity of similarity semantics in complicated networks, our proposed framework, \HINSim, accepts user-defined meta-paths as guidance to learn vertex vectors in a user-preferred embedding space.
Moreover, an efficient and parallel sampling-based optimization algorithm has been developed to learn the embeddings of HINs.
Extensive experiments on real-world large-scale HINs demonstrate a significant improvement on the effectiveness of \HINSim over several state-of-the-art algorithms and its efficiency. 
}
\end{abstract}

\section{Introduction}

A \emph{heterogeneous information network} (\emph{HIN}) is a network that consists of multi-typed vertices connected via multi-typed edges.
Modeling data in the real world as heterogeneous information networks (\emph{HINs}) can capture rich data semantics and facilitate various applications~\cite{sun2011pathsim,jeh2002simrank,hallac2015network,tang2015line,shang2014parallel,Sun:2013:MHI:2481244.2481248}.
\emph{Similarity search} in HINs is a fundamental problem for mining large HINs and much digital ink has been spilled over it in the community (e.g., \cite{sun2011pathsim,Sun:2013:MHI:2481244.2481248,tang2015line,tang2015pte}).
In this paper, we are particularly interested in utilizing HIN to conduct similarity search among the objects of the same type.
For example, given a social media network in Yelp with connections between reviews, users and businesses, we can find similar restaurants (i.e., similarity search among businesses), and recommend potential friends with similar preferences (i.e., similarity search among users).

Naturally, people tend to employ different semantics in their network-based search even towards the same application.
Take a bibliographic network as an example that consists of papers, authors, venues, and terms.
To find similar authors to a given author, some users may weigh more on shared technical terms in papers, while others more on shared publication venues.
\PathSim~\cite{sun2011pathsim} proposes to use \emph{meta-paths} to define and guide similarity search in HIN, with good success.
A meta-path is represented by a path (i.e., a connected sequence of vertices) at the schema level (e.g., $\langle$author$-$paper$-$author$\rangle$).
However, \PathSim has not explored similarities embedded deeply in the HIN structure.
When hunting for similar authors, if the given meta-path is $\langle$author$-$paper$-$venue$-$paper$-$author$\rangle$, \PathSim can only build bridges for authors publishing papers in the same venues (e.g., ``WSDM''), but cannot efficiently explore the semantic similarity between venues (e.g., ``WSDM'' and ``WWW'') to improve the search.
However, such kind of semantic similarity can be easily implied by embedded semantic structure of the HIN.
For an extreme example, if one author only publish in ``WSDM'' while the other only has publications in ``WWW'', their \PathSim similarity will be 0.
Although a bridge can be built between the two similar venues by traversing some long paths,
it becomes much more costly and does not seem to be an elegant way compared with the embedding-based approach studied in this paper.
\nop{ 
Moreover, \PathSim doesn't have the embedding vectors of vertices, which can make the further analysis more efficient, such as clustering.
}

Along another line of study, network-embedding techniques have been recently explored for homogeneous information networks, which treat all vertices and edges as of the same type, represented by \LINE~\cite{tang2015line}.
An alternative and better way is to first project the HIN to several bipartite (assuming user-given meta-paths are symmetric) or homogeneous networks and then apply the edge-wise HIN embedding technique, such as \PTE~\cite{tang2015pte}.
However, the network projection itself is count-based which does not preserve the underlying semantics, and the cost for computing such projected networks is high.
For example, taking a user-selected meta-path $\langle$author$-$paper$-$venue$-$paper$-$author$\rangle$ (i.e., shared-venues) to find similar authors,
the projection will generate a homogeneous network consisting of only authors, but it loses important network structure information (e.g., venues and papers) ~\cite{Sun:2013:MHI:2481244.2481248} and leads to a rather densely connected network (since many authors may publish in many venues).
Clearly, direct modeling of the original heterogeneous information network will capture richer semantics than exploiting the projected networks.

Acknowledging the deficiency of the above two types of approaches, we propose \HINSim, a novel embedding-based similarity search framework, with the following contributions.
\begin{itemize}[noitemsep,nolistsep]
\parskip -0.2ex
\item A general embedding-based similarity search framework is proposed for HINs, where an HIN may contain undirected, directed, weighted, and un-weighted edges as well as various types of vertices;
\item Our framework incorporates a set of meta-paths as an input from a user to better capture the semantic meaning of user-preferred similarity; and
\item It handles large-scale HINs efficiently due to a novel sampling method and a parallel optimization framework.
\end{itemize}

To the best of our knowledge, this is the first work that proposes a general meta-path guided embedding framework for similarity search in heterogeneous information networks.
\vspace{-0.2cm}
\section{Related Work}\label{sec:rel}
\vspace{-0.2cm}
\subsection{Meta-Path Guided Similarity Search}
The concept of meta-path, which represents a connected sequence of vertices at the schema level, plays a crucial role in typed and structured search and mining in HINs.
\PathSim~\cite{sun2011pathsim} defines the similarity between two vertices of the same type by the normalized count of path instances following a user-specified meta-path between any pair of vertices.
\cite{sun2011pathsim} shows that the \PathSim measure captures better peer similarity semantics than random walk-based similarity measures, such as P-PageRank~\cite{jeh2003scaling} and SimRank~\cite{jeh2002simrank}. Moreover, \cite{sun2012integrating} shows that user guidance can be transformed to a weighted combination of meta-paths.
However, \PathSim does not explore the similarity \emph{embedded} in the structure of a HIN.
Moreover, \PathSim doesn't have the embedding vectors of vertices, which can make the further analysis more efficient, such as clustering.

\vspace{-0.1cm}
\subsection{Embedding-based Similarity Search}
Recently, embedding technique, which aims at learning low-dimensional vector representations for entities while preserving proximities, has received an increasing attention due to its great performance in many different types of tasks.
As a special and concrete scenario, embedding of homogeneous networks containing vertices of the same type has been studied recently.
\LINE~\cite{tang2015line} and {\sf DeepWalk}~\cite{perozzi2014deepwalk} utilize the network link information to construct latent vectors for vertex classification and link prediction.
{\sf DCA}~\cite{cho2015diffusion} starts from the personalized PageRank but does further decomposition to get better protein-protein interaction predictions in biology networks.
However, these homogeneous models cannot capture the information about entity types nor about relations across different typed entities in HINs.

There are also embedding algorithms developed for HINs.
For example, Chang et al.\ propose to incorporate deep neural networks to train embedding vectors for both text and images at the same time~\cite{chang2015heterogeneous}.
Under a supervised setting, \PTE~\cite{tang2015pte} utilizes labels of words and constructs bipartite HINs to learn predictive embedding vectors for words.
Embedding techniques have been also applied to knowledge graphs to resolve question-answering tasks~\cite{gu2015traversing} and retain knowledge relations between entities~\cite{xie2016representation}.
However, these are all specially designed for specific types of networks and tasks and thus difficult to be extended to incorporate user guidance.
The vector spaces constructed by different methods have different semantic meanings due to the statistics they emphasize.
In many real-world scenarios, it is often difficult to find an appropriate model, and the models have to be revised to fit into the desired usage.

\vspace{-0.1cm}
\subsection{Similarity Search in Vector Spaces}
Various similarity metrics have been designed for vector spaces in different tasks, such as cosine similarity~\cite{singhal2001modern}, Jaccard coefficient~\cite{levandowsky1971distance}, and the $p$-norm distance~\cite{duren1970theory}.
In our work, we directly utilize cosine similarity based on embedding vectors of vertices.
Many efforts have been paid on optimizing the efficiency of top-$k$ nearest neighbor search~\cite{yianilos1993data,katayama1997sr,indyk1998approximate,muja2009fast,andoni2006near}.
We adopt these existing efficient similarity search techniques to support online queries.


\section{Preliminaries}\label{sec:pre}

In this section, a series of definitions and notations are presented.

\vspace{-1mm}
\begin{thm:def}
\textbf{Heterogeneous Information Network} is an information network where both vertices and edges have been associated with different types.
In a heterogeneous information network $G=(V, E, R)$, $V$ is the set of typed vertices (i.e., each vertex has its own type), $R$ is the set of edge types in the network, and $E \subset V \times V \times R$ is the edge set.
An \textbf{edge} in a heterogeneous information network is an ordered triplet $e = \langle u, v, r \rangle$, where $u$ and $v$ are two typed vertices associated with this edge and $r$ is the edge type.
\end{thm:def}
In a general heterogeneous information network, there might be multiple typed edges between the same two vertices and the edge could be either directed or undirected.
The definition above naturally supports all these cases.
To better explain it, we use a bibliographic network as an example.

\begin{figure}[t]
  \centering
  \includegraphics[width=0.4 \textwidth]{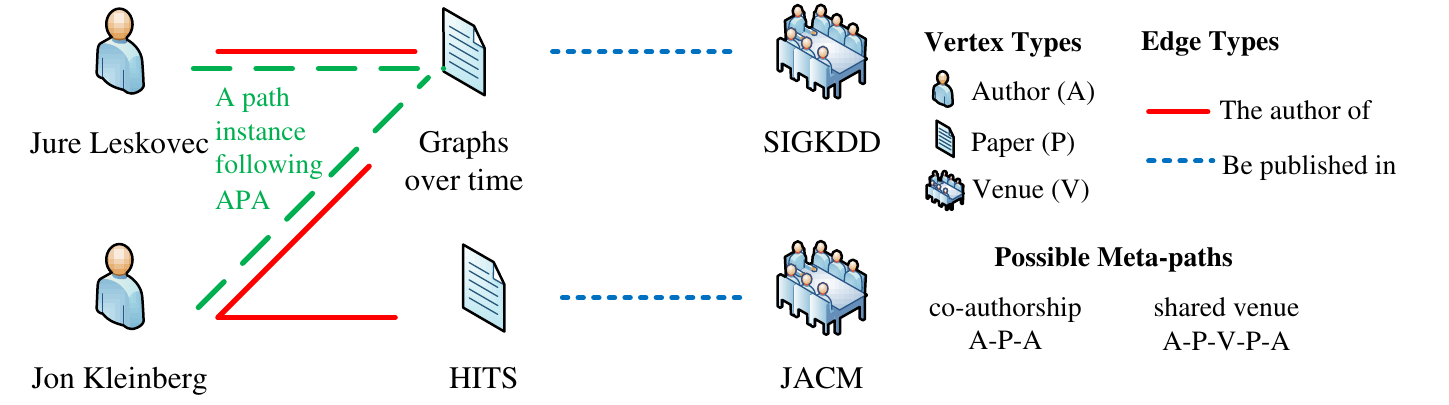}
  \vspace{-0.3cm}
  \caption{A real small bibliography network for illustrations of the definitions.}
  \label{fig:hinexample}
  \vspace{-0.3cm}
\end{figure}

\begin{thm:eg}
In the bibliographic network as shown in Fig.~\ref{fig:hinexample}, the vertex set $V$ consists of three types, $\{$\mquote{author}, \mquote{paper}, \mquote{venue}$\}$, the edge type set $R$ contains \mquote{the author of the paper} and \mquote{the paper was published in the venue}.
The edge set $E$ consists of concrete edges like $\langle u =$ \mquote{Jon Kleinberg}, $v =$ \mquote{HITS}, $r = $ \mquote{the author of} $\rangle $, where \emph{Jon Kleinberg} is one of the authors of the paper \emph{HITS}.
For ease of representation, edge type $r$ is denoted as $A$$-$$P$ once there is only one edge type between author and paper.
Another edge is $\langle u =$ \mquote{HITS}, $v =$ \mquote{JACM}, $r = $ \mquote{be published in} $\rangle $, which means the paper was published in the venue \mquote{JACM}.
In this case, $r$ is denoted as $P$$-$$V$.
Note that both edge types are undirected here.
That is, $A$$-$$P$ and $P$$-$$A$ are actually referring to the same edge type.
So as $P$$-$$V$ and $V$$-$$P$.
\end{thm:eg}

We define the concepts of meta-path and sub-meta-path by concatenating edge types in a sequence, as follows.

\begin{thm:def}
In a HIN $G=(V, E, R)$, a \textbf{meta-path} is a sequence of compatible edge types $\mathcal{M}=\langle r_1, r_2, \ldots, r_L \rangle$ with length $L$, $\forall r_i \in R$, and the outgoing vertex type of $r_i$ should match the incoming vertex type of $r_{i + 1}$. For any $1 \le s \le t \le L$,  we can induce a \textbf{sub-meta-path} $\mathcal{M}_{s, t} = \langle r_s, r_{s + 1}, \ldots, r_t \rangle$.
\end{thm:def}
Particularly, an edge type $r$ can be viewed as a length-$1$ meta-path $\mathcal{M}=\langle r \rangle$.

A sequence of edges following $\mathcal{M}$ is called a path instance.
Because there might be multiple edges between the same pair of vertices, instead of the vertex sequence, the edge sequence is used to describe a path instance.
Formally speaking,
\begin{thm:def}
Given a HIN $G=(V, E, R)$ and a meta-path $\mathcal{M}=\langle r_1, r_2, \ldots, r_L \rangle$, any path $\mathcal{P}_{e_1\rightsquigarrow e_L} = \langle e_1, e_2, \ldots, e_L \rangle$ connecting vertices $u_1$ and $u_{L+1}$ (i.e., $v_L$), is a \textbf{path instance} following $\mathcal{M}$, if and only if $\forall 1 \le i \le L$, the $i$-th edge is type-$r_i$, and $\forall 1 \le i \le L, v_i = u_{i+1}$.
\end{thm:def}

We continue the example of the bibliographic network and list some meta-paths and path instances.
\begin{thm:eg}
In the bibliographic network as shown in Fig.~\ref{fig:hinexample}, a length-$2$ meta-path $\langle A$$-$$P, P$$-$$A \rangle$ (abbrev.\ as $A$$-$$P$$-$$A$) expresses the co-authorship relation. The collaboration between \mquote{Jure Leskovec} and \mquote{Jon Kleinberg} on the paper \mquote{Graphs over time} is a path instance of $A$$-$$P$$-$$A$. Similarly, a length-$4$ meta-path $A$$-$$P$$-$$V$$-$$P$$-$$A$ captures the shared venues and any two authors published a certain paper in a same venue could be its path instance. Besides, $P$$-$$V$$-$$P$ is a sub-meta-path of $A$$-$$P$$-$$V$$-$$P$$-$$A$.
\end{thm:eg}

\begin{thm:def}
\textbf{Meta-path Guided Similarity Search} is a similarity search task on HINs, where the semantic meanings of the similarity are determined by $n$ meta-paths $\{ \mathcal{M}_1, \mathcal{M}_2, \ldots, \mathcal{M}_n \}$ specified by the user.
\end{thm:def}

\begin{thm:eg}
In the bibliographic network, to find similar authors, a user may choose two meta-paths $A$$-$$P$$-$$V$$-$$P$$-$$A$ and $A$$-$$P$$-$$A$ as guidance.
\end{thm:eg}

In the subsequent discussion, we focus on the case of a single meta-path $\mathcal{M}$.
This is because (1) the principal ideas for exploring multiple weight-assigned meta-paths are essentially the same; (2) in our experiments, the performance gain of the optimal weighted combination of meta-paths is not significant.
Moreover, we leave the study on derivation of a weighted combination of multiple meta-paths based on user's high-level guidance (i.e., user providing examples instead of path weights explicitly) to future work.
Such a user-guided approach without adopting the embedding framework has been studied in \cite{yu2012user,sun2012integrating,meng2015discovering}.


\section{Methodology}\label{sec:method}
In this section, to incorporate meta-paths, we formulate a probabilistic embedding model inspired from many previous studies.
We propose an efficient and scalable optimization algorithm relying on the sampling of path instances following the given meta-path.
Our proposed efficient sampling methods are the most crucial steps and thus are discussed separately.
In addition, we also provide thorough time complexity analysis.

\subsection{Probabilistic Embedding Model Incorporating Meta-Paths}

\noindent\textbf{Model Formulation}. The basic idea of our approach is to preserve the HIN structure information into the learned embeddings, such that vertices which co-occur in many path instances turn to have similar embeddings.

To preserve the structure of a HIN $G=(V, E, R)$, we first define the conditional probability of vertex $v$ connected to vertex $u$ by some path instances following the meta-path $\mathcal{M}$ as:
    \begin{equation}\label{eq:prob_edge}
    	Pr(v|u, \mathcal{M}) = \frac{\exp(f(u,v,\mathcal{M}))}{\sum_{v' \in V}\exp(f(u,v',\mathcal{M}))}
    \end{equation}
    where function $f$ is a scoring function modeling the relevance between $u$ and $v$ conditioned on the meta-path $\mathcal{M}$. In particular, we encode the meta-path through the following formulation inspired from~\cite{pennington2014glove,shang2014parallel}:
    \begin{equation*}
    	f(u,v,\mathcal{M}) = \mu_\mathcal{M} + \mathbf{p_\mathcal{M}}^T\mathbf{x_u} + \mathbf{q_\mathcal{M}}^T\mathbf{x_v} + \mathbf{x_u}^T \mathbf{x_v}
    \end{equation*}
    Here, $\mu_{\mathcal{M}} \in \mathbb{R}$ is the global bias of the meta-path $\mathcal{M}$, $\mathbf{p_\mathcal{M}}$ and $\mathbf{q_\mathcal{M}} \in \mathbb{R}^d$ are local bias vectors which are $d$ dimensional. $\mathbf{x_u}$ and $\mathbf{x_v} \in \mathbb{R}^d$ are $d$ dimensional embedding vectors for vertices $u$ and $v$ respectively. Under such definition, if the embeddings of two vertices have a larger dot product, the two vertices are likely having a larger relevance score, and thus co-occuring in many path instances. Note that if users want a symmetric score function, $\forall u, v, f(u, v, \mathcal{M}) = f(v, u, \mathcal{M})$, we can restrict $\mathbf{p_\mathcal{M}} = \mathbf{q_\mathcal{M}}$.

    For a better understanding of the scoring function $f$, we can rewrite it as follows
    \begin{equation*}
    	f(u,v,\mathcal{M}) = (\mu_\mathcal{M} - \mathbf{p_\mathcal{M}}^T\mathbf{q_\mathcal{M}})  + (\mathbf{x_u} + \mathbf{q_\mathcal{M}})^T(\mathbf{x_v} + \mathbf{p_\mathcal{M}})
    \end{equation*}
    where we can see that $\mathbf{p_\mathcal{M}}$ and $\mathbf{q_\mathcal{M}}$ shift $\mathbf{x_u}$ and $\mathbf{x_v}$ according to the semantic of the meta-path $\mathcal{M}$ while $\mu_\mathcal{M}$ adjusts the score to an appropriate range.

For a path instance $\mathcal{P}_{e_1\rightsquigarrow e_L} = \langle e_1 = \langle u_1, v_1, r_1\rangle , e_2 = \langle u_2, v_2, r_2\rangle , \ldots, e_L = \langle u_L, v_L, r_L\rangle \rangle$ following the meta-path $\mathcal{M}=\langle r_1, r_2, \ldots, r_L \rangle$, we adopt the following approximation, by approximating the probability of the first vertex.
    \begin{equation}\label{eq:path}
    \begin{split}
    	Pr(\mathcal{P}_{e_1\rightsquigarrow e_L}|\mathcal{M}) =&\ Pr(u_1|\mathcal{M}) \times Pr(\mathcal{P}_{e_1\rightsquigarrow e_L}|u_1,\mathcal{M})\\
    	\propto &\ C(u_1, 1|\mathcal{M})^{\gamma} \times Pr(\mathcal{P}_{e_1\rightsquigarrow e_L} | u_1,\mathcal{M})
    \end{split}
    \end{equation}
    where $C(u, i |\mathcal{M})$ represents the number of path instances following $\mathcal{M}$ with the $i^{th}$ vertex being $u$. $\gamma$ is a widely used parameter to control the effect of overly-popular vertices, which is usually $3/4$ inspired from~\cite{mikolov2013distributed}. In Sec.~\ref{sec:dp}, we will show an efficient dynamic programming algorithm to compute $C(u, i|\mathcal{M})$.

The conditional probability, $Pr(\mathcal{P}_{e_1\rightsquigarrow e_L} | u_1,\mathcal{M})$, is now the last undefined term. The simplest definition is $Pr(v_L | u_1,\mathcal{M})$, which assumes that the probability only depends on the two ends of the path instance and directly applies Eq.~(\ref{eq:prob_edge}). However, it omits the intermediate information and is equivalent to projection-based models.

In this paper, we propose two possible solutions as follows, and later show ``pairwise'' is more effective than ``sequential'', since it exploits the meta-path guidance in a more thorough way.
    \begin{itemize}[noitemsep,nolistsep]
        \item \textbf{Sequential} (seq): In this setting, we assume that a vertex is highly relevant to its left/right neighbors in the sequence: $Pr(\mathcal{P}_{e_1\rightsquigarrow e_L} | u_1,\mathcal{M}) = \prod_{k=1}^{L} Pr(v_k | u_k,\mathcal{M}_{k,k})$. 
        \item \textbf{Pairwise} (pair): In this setting, we assume all vertices in a path instance are highly relevant to each other, and thus the probability of the path instance is defined as
        $Pr(\mathcal{P}_{e_1\rightsquigarrow e_L} | u_1,\mathcal{M}) = \prod_{s=1}^{L}\prod_{t=s}^{L} Pr(v_t | u_s,\mathcal{M}_{s, t})$. 
        As a result, vertices co-occur in many path instances turn to have large relevance scores.
    \end{itemize}

\noindent\textbf{Noise-Contrastive Estimation (NCE)}.
Given the conditional distribution defined in Eqs.~(\ref{eq:prob_edge}) and (\ref{eq:path}), the maximum likelihood training is tractable but expensive because computing the gradient of log-likelihood takes time linear in the number of vertices.

Since learning rich vertex embeddings does not require the accurate probability of each path instance, we adopt the NCE for optimization, which has been proved to significantly accelerate the training without cutting down the embedding qualities~\cite{gutmann2012noise,Mnih12afast}. The basic idea is to sample some observed path instances associated with some noisy ones, and it tries to maximize the probability of each observed instance while minimize the probability of each noisy one.

	Specifically, we reduce the problem of density estimation to a binary classification, discriminating between samples from path instances following the user selected meta-path and samples from a known noise distribution.
    In particular, we assume these samples come from the mixture.
    \begin{equation*}
    	\frac{1}{K+1}Pr^{+}(\mathcal{P}_{e_1\rightsquigarrow e_L}|\mathcal{M}) + \frac{K}{K+1}Pr^{-}(\mathcal{P}_{e_1\rightsquigarrow e_L}|\mathcal{M})
    \end{equation*}
    where {\scriptsize $Pr^{+}(\mathcal{P}_{e_1\rightsquigarrow e_L}|\mathcal{M})$} denotes the distribution of path instances in the HIN following the meta-path $\mathcal{M}$. {\scriptsize $Pr^{-}(\mathcal{P}_{e_1\rightsquigarrow e_L}|\mathcal{M})$} is a noise distribution, and for simplicity we set
    \begin{equation*}
    	Pr^{-}(\mathcal{P}_{e_1\rightsquigarrow e_L}|\mathcal{M}) \propto \prod_{i = 1}^{L+1} C(u_i, i\,|\mathcal{M})^{\gamma}
    \end{equation*}
    We further assume the noise samples are $K$ times more frequent than positive path instance samples. The posterior probability that a given sample came from the path instance samples following the given meta-path is
    \begin{equation*}
    	\frac{Pr^{+}(\mathcal{P}_{e_1\rightsquigarrow e_L}|\mathcal{M})}{Pr^{+}(\mathcal{P}_{e_1\rightsquigarrow e_L}| \mathcal{M}) + K \cdot Pr^{-}(\mathcal{P}_{e_1\rightsquigarrow e_L}|\mathcal{M})}
    \end{equation*}

    Since we would like to fit {\scriptsize $Pr(\mathcal{P}_{e_1\rightsquigarrow e_L}|\mathcal{M})$} to {\scriptsize $Pr^{+}(\mathcal{P}_{e_1\rightsquigarrow e_L}|\mathcal{M})$}, we simply maximize the following expectation.
    \begin{equation*}\scriptsize
    \begin{split}
    	\mathcal{L}_{\mathcal{M}} = &\mathbb{E}_{Pr^{+}}\bigg[\log \frac{Pr(\mathcal{P}_{e_1\rightsquigarrow e_L}|\mathcal{M})}{Pr(\mathcal{P}_{e_1\rightsquigarrow e_L}|\mathcal{M}) + KPr^-(\mathcal{P}_{e_1\rightsquigarrow e_L}|\mathcal{M})} \bigg]\\
    	   &+K\, \mathbb{E}_{Pr^{-}}\bigg[\log \frac{KPr^-(\mathcal{P}_{e_1\rightsquigarrow e_L}| \mathcal{M})}{Pr(\mathcal{P}_{e_1\rightsquigarrow e_L}| \mathcal{M}) + KPr^-(\mathcal{P}_{e_1\rightsquigarrow e_L}|\mathcal{M})}  \bigg]\\
    \end{split}
    \end{equation*}

    Suppose we are using the sequential definition as Eq.~(\ref{eq:objective_sequential}). The loss function derived from NCE becomes
    \begin{equation*}\scriptsize
    \begin{split}
    	\mathcal{L}_{\mathcal{M}, \mbox{seq}} = & \mathbb{E}_{Pr^{+}}\bigg[\log \sigma( \Delta_{e_1\rightsquigarrow e_L | \mathcal{M}} ) \bigg] \\
    	& +K\, \mathbb{E}_{Pr^{-}}\bigg[\log \big(1 - \sigma( \Delta_{e_1\rightsquigarrow e_L | \mathcal{M}} ) )\big)  \bigg]
    \end{split}
    \end{equation*}
    where {\scriptsize $\Delta_{e_1\rightsquigarrow e_L | \mathcal{M}} = \sum_{i=1}^{L} f(u_i, v_i, \mathcal{M}) - \log \big(K \cdot Pr^-(\mathcal{P}_{e_1\rightsquigarrow e_L}| \mathcal{M})\big)$} and $\sigma(\cdot)$  is the sigmoid function.
    Note that when deriving the above equation, we used $\exp(f(u, v, \mathcal{M}))$ in place of $Pr(v|u,\mathcal{M})$, ignoring the normalization term in Eq.~(\ref{eq:prob_edge}).
    We can do this because the NCE objective encourages the model to be approximately normalized and recovers a perfectly normalized model if the model class contains the data distribution \cite{gutmann2012noise,Mnih12afast}.
    The above expectation is also studied in \cite{mikolov2013distributed}, which replaces $\Delta_{e_1\rightsquigarrow e_L | \mathcal{M}}$ with  $\sum_{i=1}^{L} f(u_i, v_i, \mathcal{M})$ for ease of computation and names the method \emph{negative sampling}.
    We follow this idea and have the approximation as follows.
    \begin{equation}\scriptsize\label{eq:objective_sequential}
    \begin{split}
    	& \mathcal{L}_{\mathcal{M}, \mbox{seq}} \approx \sum_{\mathcal{P}_{e_1\rightsquigarrow e_L} \mbox{ following } \mathcal{M}}
    	\quad \log \sigma(\sum_{i=1}^{L}f(u_i, v_i, \mathcal{M}_{i, i})) \,\,+ \\
    	 &  \textstyle\sum_{k = 1}^K \, \mathbb{E}_{\mathcal{P}^k_{e_1\rightsquigarrow e_L} \sim Pr^{-}|u_1,\mathcal{M}}\bigg[\log \big(1 - \sigma(\sum_{i=1}^{L} f(u^k_i, v^k_i, \mathcal{M}_{i, i}) )\big)  \bigg]
    \end{split}
    \end{equation}
    The following loss function under the pairwise setting can be derived from NCE utilizing the same approximation.
    \begin{equation}\scriptsize\label{eq:objective_pairwise}
    \begin{split}
    	& \mathcal{L}_{\mathcal{M}, \mbox{pair}} \approx \sum_{\mathcal{P}_{e_1\rightsquigarrow e_L} \mbox{ following } \mathcal{M}}
    	\quad \log \sigma(\sum_{i=1}^{L}\sum_{j=i}^{L} f(u_i, v_j, \mathcal{M}_{i, j})) \,\,+ \\
    	 &  \textstyle\sum_{k = 1}^K \, \mathbb{E}_{\mathcal{P}^k_{e_1\rightsquigarrow e_L} \sim Pr^{-}|u_1,\mathcal{M}}\bigg[\log \big(1 - \sigma(\sum_{i=1}^{L}\sum_{j=i}^{L} f(u^k_i, v^k_j, \mathcal{M}_{i, j}) )\big)  \bigg]
    \end{split}
    \end{equation}
	
\noindent\textbf{Online Similarity Search}.
    For any interested pairs of vertices $u$ and $v$, their similarity is defined by the cosine similarity between $\mathbf{x_u}$ and $\mathbf{x_v}$, i.e., $sim(u, v) = \frac{\mathbf{x_u}^T \mathbf{x_v}}{\|\mathbf{x_u}\|\cdot \|\mathbf{x_v}\|}$. We choose the cosine similarity metric instead of the function $f$ in Eq.~(\ref{eq:prob_edge}) because the norm of vectors $\mathbf{x_u}$ and $\mathbf{x_v}$ do not help the similarity search task~\cite{schakel2015measuring}. Moreover, cosine similarity is equivalent to Euclidean distance when $||\mathbf{x_u}|| = ||\mathbf{x_v}|| = 1$,
    which makes the top-$k$ similar vertices of the given vertex $u$ able to be efficiently solved using approximate nearest neighbors search~\cite{muja2009fast} after normalizations.

\subsection{Optimization Algorithm}
    \SetAlgoSkip{}
    \begin{algorithm}[t]
        \caption{\HINSim Training Framework}\label{alg:framework_training}
        \textbf{Require}: HIN $G=(V, E, R)$, a user-specified meta-path $\mathcal{M}$, sampling times $t$, and negative sampling ratio $K$\\
        \textbf{Return}: Vertex Embedding Vectors $\mathbf{x}_u, \forall u$\\
        initialize parameters $\mu, \mathbf{p}_{\cdot}, \mathbf{q}_{\cdot}, \mathbf{x}_{\cdot}$ \\
        \While {not converge} {
            \For {$i = 1$ {\bfseries to} $t$} {
            	 $p^{+} \leftarrow$ a sampled positive path instance following the meta-path $\mathcal{M}$\\
                Optimize for a path instance $p^{+}$ with label $1$. \\
                $s \leftarrow $ the first vertex on $p^{+}$ \\
                \For{$k = 1$ {\bfseries to} $K$} {
                    $p^{-} \leftarrow$ a sampled negative path instance following the meta-path $\mathcal{M}$ starting from $s$\\
                    Optimize for a path instance $p^{-}$ with label $0$. \\
                }
            }
        }
        \textbf{return} $\mathbf{x}_{\cdot}$
    \end{algorithm}

    The number of vertex pairs $\langle u, v \rangle$ that are connected by some path instances following at least one of user-specified meta-paths can be $O(|V|^2)$ in the worst case, which is too large for storage or processing when $|V|$ is at the level of millions and even billions, and thus makes it impossible to directly handle the projected networks over the meta-paths.
    Therefore, sampling a subset of path instances according to their distribution becomes the best and most feasible choice when optimizing, instead of going through every path instance per iteration.
    Thus even if the network itself contains a large number of edges, our method is still very efficient.
    The details of our training framework is shown in Algorithm~\ref{alg:framework_training}.

    Once a path instance following the meta-path $\mathcal{M}$ has been sampled, the gradient decent method is used to update the parameters $\mathbf{x_u}, \mathbf{x_v}, \mathbf{p_\mathcal{M}}, \mathbf{q_\mathcal{M}}$, and $\mu_\mathcal{M}$ one by one. As a result, our sampling-based training framework (Algorithm~\ref{alg:framework_training}) becomes a stochastic gradient decent framework.
    The derivations of these gradients are easy and thus are omitted.
    Moreover, many prior studies have shown that the stochastic gradient descent can be parallelized without any locks.
    For example, Hogwild~\cite{recht2011hogwild} provides a general and lock-free strategy for fully parallelizing any stochastic gradient descent algorithms in a shared memory.
    We utilize this technique to speed up our optimization via multi-threads.

    \subsubsection{Efficient Sampling}\label{sec:dp}
        Given a length-$L$ meta-path $\mathcal{M} = \langle r_1, r_2, \ldots, r_L \rangle$, there might be $O(|V|^L)$ different path instances in total. It becomes an obstacle for storing all the path instances while directly sampling over them takes a constant time.
        We propose to run a dynamic programming algorithm computing auxiliary numbers so that the online sampling part can be done in a constant time.

        \noindent\textbf{Pre-computation}.
            As mentioned in Sec.~\ref{sec:pre}, the probability of sampling a path instance following the meta-path $\mathcal{M}$ is only related to $C(u, i | \mathcal{M})$, which represents the count of path instances following the meta-path $\mathcal{M}$ with the $i^{th}$ vertex being $u$.

            First, we need to figure out the boundary case. When $i = L + 1$, for any vertex $u$, if it is possible to be the next vertex in an edge of $r_L$ (i.e., it could be $v_L$), we have $C(u, L + 1 | \mathcal{M}) = 1$. Otherwise, it should be $0$.

            Then, we derive the following recursion formula when $1 \le i \le L$ for any vertex $u$.
            \begin{equation}
                C(u, i | \mathcal{M})  = \sum_{v | \langle u, v, r_i\rangle \in E} C(v, i + 1 | \mathcal{M})
            \end{equation}

            An efficient way to do the summation in this formula is to traverse all type-$r_i$ edges starting from vertex $u$ as shown in Algorithm~\ref{alg:prep_c}. Its detailed time complex analysis will be presented later.

            \SetAlgoSkip{}
            \begin{algorithm}[t]
                \caption{Pre-computation of $C(u, i | \mathcal{M})$}\label{alg:prep_c}
                \textbf{Require}: HIN $G=(V, E, R)$ and meta-path $\mathcal{M} = \langle r_1, r_2, \ldots, r_L \rangle$ \\
                \textbf{Return}: $C(u, i | \mathcal{M})$ \\
                \tcc{\small \textcolor{blue}{initialization}}
                \For {each vertex $u \in V$} {
                    \If {$u$ is possibly as the second vertex in $r_L$} {
                        $C(u, L + 1 | \mathcal{M}) \leftarrow 1$ \\
                    }
                    \Else {
                        $C(u, L + 1 | \mathcal{M}) \leftarrow 0$ \\
                    }
                }
                \tcc{\small \textcolor{blue}{dynamic programming}}
                \For {$i \leftarrow L$ {\bfseries to} $1$} {
                    \For {each vertex $u \in V$} {
                        $C(u, i | \mathcal{M}) \leftarrow 0$ \\
                    }
                    \For {each type $r_i$ edge $\langle u, v, r_i\rangle$} {
                        $C(u, i | \mathcal{M}) \leftarrow C(u, i | \mathcal{M}) + C(v, i + 1 | R)$ \\
                    }
                }
                \textbf{Return}: $C(u, i | \mathcal{M})$ \\
            \end{algorithm}

        \noindent\textbf{Online Sampling}.
            Based on the pre-computed $C(u, i | \mathcal{M})$, one can easily figure out an efficient online sampling method for path instances following the user-given meta-path $\mathcal{M}$.
            The key idea is to sample the vertices on the path instance one by one.
            That is, the $i$-th vertex is conditioned on the previous $i-1$ vertices.
            As shown in Algorithm~\ref{alg:sample_postive_p}, the sampling pool for the $i$-th vertex is restricted to the adjacent vertices (via type-$r_i$ edges) of the previous $(i-1)$-th vertex.

            However, things are a little different when dealing with the negative path instances.
            First, the negative path instances are associated with a positive path instance and thus the first vertex is fixed.
            Second, the remaining vertices on the negative path instances are independent.
            Therefore, they are all sampled from $V$ based on $\propto C(u, i | \mathcal{M})^{\gamma}$, $\forall i > 1$.
%

\subsubsection{Weighted Combination}

Sometimes, due to the subtle semantic meanings of the similarity, instead of a single meta-path, the weighted combination of $n$ meta-paths could enhance the performance of similarity search. Suppose $\{\lambda_1, \lambda_2, \ldots, \lambda_n\}$ are the weights ($\forall i, \lambda_i > 0$ and $\sum_{i=1}^{n} \lambda_i = 1$), the unified loss function becomes the weighted sum over the loss functions of individual meta-paths based on the weights. That is, $\mathcal{L}_{\mbox{seq}} = \sum_{i = 1}^{n} \lambda_i \mathcal{L}_{\mathcal{M}_i, \mbox{seq}}$ and $\mathcal{L}_{\mbox{pair}} = \sum_{i = 1}^{n} \lambda_i \mathcal{L}_{\mathcal{M}_i, \mbox{pair}}$. The Algorithm~\ref{alg:framework_training} can be modified accordingly by first sampling a meta-path $\mathcal{M}$ from $\forall j, Pr(\mathcal{M}_j) = \lambda_j$ in the beginning of the ``while'' loop.

The weighted combination of meta-paths can be either explicitly specified by users or learned from a set of similar/dissimilar examples provided by users.
Such user-guided meta-path generation has been studied in \cite{sun2012integrating} without considering embedding.
Because weight learning is beyond the scope of this paper, we leave such an extension to embedding-based similarity search to a future work, and adopt grid searches to obtain the best weighted combination of meta-paths in Sec.~\ref{sec:auc} assuming we have the groundtruth.

            \SetAlgoSkip{}
            \begin{algorithm}[t]
                \caption{Sample a positive path instance $p^{+}$}\label{alg:sample_postive_p}
                \textbf{Require}: HIN $G=(V, E, R)$, meta-path $\mathcal{M} = \langle r_1, r_2, \ldots, r_L \rangle$, $C(u, i | \mathcal{M})$, and weighting factor $\gamma$ \\
                \textbf{Return}: a positive path instance following $\mathcal{M}$ \\
                $u_1 \leftarrow $ a random vertex $ \propto C(u_1, 1 | \mathcal{M})^{\gamma}$ from $V$ \\
                \For {$i = 1$ {\bfseries to} $L$} {
                    $V_i \leftarrow \{v | \langle u_i, v\rangle \in E_{r_i}\}$ \\
                    \tcc{\small \textcolor{blue}{$\gamma$ is only applied at the first vertex when sampling positive path instances.}}
                    $v_i \leftarrow$ a random vertex $\propto C(u_i, i | \mathcal{M})$ from $V_i$\\
                    \If {$i <  L$} {
                        $u_{i + 1} \leftarrow v_i$
                    }
                }
                \textbf{return} $\mathcal{P}_{e_1\rightsquigarrow e_L} = \langle e_1 = \langle u_1, v_1, r_1\rangle, e_2 = \langle u_2, v_2, r_2\rangle, \ldots, e_L = \langle u_L, v_L, r_L\rangle \rangle$
            \end{algorithm}

\subsubsection{Complexity Analysis}

\noindent \textbf{Pre-computation.}  For a given length-$L$ meta-path $\mathcal{M}$, we have $u \in V$ and $1 \le i \le L + 1$, which means the memory complexity is $O(|V|L)$.
For each given $i$, we only have to consider all type-$r_i$ edges starting from different vertices, which implies the time complexity is $O((|V| + |E|) L)$.

\smallskip
\noindent \textbf{Online sampling.}
We have adopted the alias method~\cite{walker1977efficient} to make the online sampling from any discrete distribution $O(1)$.
Therefore, we have to precompute all discrete distributions and restore them in the memory.
Given a length-$L$ mete-path $\mathcal{M}$, for the negative sampling, we have $O(L)$ different distributions and each of them is over $|V|$ discrete values; for the positive sampling there are $O(|V| L)$ different distributions but the number of variables depends on the number of edges of the certain type.
The total discrete values they have is $O(|E| L)$. In summary, both the time and memory complexities of the preparation of the alias method are $O((|V|+|E|) L)$, while the time complexity of sampling any path instance becomes $O(L)$.

\smallskip
\noindent \textbf{Optimization for a path instance.}
For a specific path instance $\mathcal{P}_{e_1\rightsquigarrow e_L}$, the time complexity is $O(dL)$, and $O(dL^2)$ for different loss functions $\mathcal{L}_{\mbox{seq}}$ and $\mathcal{L}_{\mbox{pair}}$ respectively.

\smallskip
\noindent \textbf{Overall.}
In our framework, for Algorithm~\ref{alg:framework_training}, there are $O(tK)$ times of path optimization per iteration.
Suppose there are $T$ iterations before convergence, considering the choices of different loss functions, $\mathcal{L}_{\mbox{seq}}$ and $\mathcal{L}_{\mbox{pair}}$, the overall time complexity is $O(TtKLd)$ and $O(TtKL^2d)$ respectively.
In addition, considering the complicated case of $n$ user-specified meta-paths, $O(n(|V| + |E|)L)$ has to be paid for pre-computations before sampling, where $L$ is the maximum length of given meta-paths.

\smallskip
\noindent \textbf{Parallelization.}  It has been proved that stochastic gradient descent can be fully parallelized without locks~\cite{recht2011hogwild}.
If we use $k$ threads, although the pre-computations remain the same, the time complexity of training can be $k$ times faster.

\section{Experiments}\label{sec:exp}

In this section, we evaluate our proposed model \HINSim, comparing with several state-of-the-art methods on two real-world large-scale HINs both quantitatively and qualitatively.
Also, we evaluate the sensitivity of parameters and the efficiency of \HINSim.

\subsection{Datasets}
    \begin{table}[t]
        \center
        \caption{Dataset Statistics. Whole networks in DBLP and Yelp are utilized in training, while grouping labels are only used for evaluation.}
        \vspace{-0.2cm}
        \label{tbl:dataset}
        \scalebox{0.7}{
        \begin{tabular}{|c|c|c|}
        \hline
        Dataset & DBLP & Yelp \\
        \hline
        \hline
        Node Types & paper(P), author(A),  & review(R), name(N), \\
                   & term(T), venue(V)     & business(B), word(W) \\
        \hline
        Edge Types & $P$$-$$A$, $P$$-$$T$, $P$$-$$V$ & $B$$-$$N$, $R$$-$$B$, $R$$-$$W$\\
        \hline
        \# of Vertices & 2,762,595 & 1,616,341 \\
        \hline
        \# of Edges &103,059,616  & 76,708,201 \\
        \hline
        Interesting Meta-paths & $A$$-$$P$$-$$A$, $A$$-$$P$$-$$V$$-$$P$$-$$A$, & $B$$-$$R$$-$$W$$-$$R$$-$$B$, \\
                               & $A$$-$$P$$-$$T$$-$$P$$-$$A$                   & $B$$-$$N$$-$$B$ \\
        \hline
        Two Grouping Labels  & Research Area/Group & Business/Restaurant Type\\
        \hline
        \end{tabular}
        }
        \vspace{-0.5cm}
    \end{table}

Table~\ref{tbl:dataset} shows the statistics of two real-world large-scale HINs: \textbf{DBLP} and \textbf{Yelp}.
The first dataset, \textit{DBLP}, is a bibliographic network in computer science, which includes papers~(P), authors~(A), venues~(V), and terms~(T) as four types of vertices and takes $P$ as the center in a star network schema.
There are 3 types of undirected edges: $P$$-$$A$, $P$$-$$V$, and $P$$-$$T$.
The interesting meta-paths that may be specified by users are the co-authorship meta-path $A$$-$$P$$-$$A$, the shared venue meta-path $A$$-$$P$$-$$V$$-$$P$$-$$A$, and the shared term meta-path $A$$-$$P$$-$$T$$-$$P$$-$$A$.
We have the following two groupings labeled by human experts.
\begin{itemize}[noitemsep,nolistsep]
    \item \textbf{Research Area.} We use the 4-area grouping in \textit{DBLP} labeled by human experts, which was used when evaluating \PathSim~\cite{sun2011pathsim}. There are 3,750 authors from 4 diverse research domains of computer science: \mquote{data mining}, \mquote{database}, \mquote{machine learning} and \mquote{information retrieval}.
    \item \textbf{Research Group.} This grouping in \textit{DBLP} is also labeled by human experts and is more fine-grained comparing to the \textit{Research Area} grouping. There are 103 authors from 4 research groups: \mquote{Christos Faloutsos}, \mquote{Jiawei Han}, \mquote{Michael I.\ Jordan}, and \mquote{Dan Roth}.
\end{itemize}

The second dataset, \textit{Yelp}, is a social media network of Yelp, released in Yelp Dataset Challenge\footnote{\url{https://www.yelp.com/academic_dataset}}.
This network includes businesses~(B), words in business names~(N), reviews of businesses~(R), and words in reviews~(W) as vertices.
There are 3 different types of undirected edges: $B$$-$$N$, $B$$-$$R$, and $R$$-$$W$.
The interesting meta-paths that may be specified by users are the shared review word meta-path $B$$-$$R$$-$$W$$-$$R$$-$$B$ and the shared name word meta-path $B$$-$$N$$-$$B$.
We have the following two groupings extracted from the meta-data provided in \textit{Yelp} dataset.
    \begin{itemize}[noitemsep,nolistsep]
    \item \textbf{Business Type.}
    There are various business types in the \textit{Yelp} dataset, such as \mquote{restaurants}, \mquote{hotels}, \mquote{shopping}, and \mquote{health and medical}.
    Businesses with multiple categories have been discarded to avoid ambiguity.
    To keep the results from being biased by some dominating types, we randomly sample $881$ businesses from each of these four types as labeled data, because the $4$-th popular type contains that many businesses.
    \item \textbf{Restaurant Type.}
    Since the majority of businesses in the \textit{Yelp} dataset are restaurants, we look deep into them by dividing them into different types of cuisines.
    More specifically, we have sampled $270$ restaurants from $5$ cuisines respectively: \mquote{Chinese}, \mquote{Japanese}, \mquote{Indian}, \mquote{Mexican}, and \mquote{Greek}.
    As a result, there are in total $1350$ labeled restaurants in our labeled dataset.
    \end{itemize}

\subsection{Experimental Setting}
\noindent\textbf{Meta-path.}
We select different meta-paths for different datasets to see how the meta-paths will reflect the user-preferred similarity and affect the performance. In addition, we run grid searches against different groupings to obtain the best weights of different meta-paths for \HINSim models.

\noindent\textbf{Compared Algorithms and Notations.}
We select the previous state-of-the-art algorithm in the meta-path guided similarity search problem, \PathSim, which has been reported to beat many other simiarity search methods, for example, SimRank~\cite{jeh2002simrank}, P-PageRank~\cite{ivan2011web}, random walk, and pairwise random walk. In addition, we also consider (heterogeneous) network embedding methods, such as \LINE and \PTE, which beat other embedding methods like graph factorization~\cite{ahmed2013distributed} and DeepWalk~\cite{perozzi2014deepwalk}. More details about these methods are as follows.
    \begin{itemize}[noitemsep,nolistsep]
    \item \PathSim~\cite{sun2011pathsim} is a meta-path guided similarity search algorithm which utilizes the normalized count of path instances following the user selected meta-path between any pair of vertices. When the meta-path involves text (e.g., $A$$-$$P$$-$$T$$-$$P$$-$$A$), \PathSim becomes a \emph{text-based similarity} --- the cosine similarity using bag-of-words.
    \item \LINE~\cite{tang2015line} is an embedding algorithm specifically designed for homogeneous networks, which considers both first and second order information in a network (i.e., the neighbors and the neighbors of the neighbors). By treating all vertices and edges in the HIN as homogeneous ones, we can directly apply \LINE and denote the model with first order only as \textbf{\LINE-1st} and the model using the second order information as \textbf{\LINE-2nd} respectively. One can also project the HIN to a weighted homogeneous network based on the user selected meta-path and apply \LINE. However, based on our experiments, the results are always worse than \PTE and thus omitted.
    \item \PTE~\cite{tang2015pte} decomposes a HIN to a set of edgewise bipartite networks and then learn embedding vectors. To adapt this method to our settings, the way with the best performance we discovered is to project the HIN to a weighted bipartite HIN based on the user selected meta-path. For example, if the selected meta-path is the shared venue meta-path $A$$-$$P$$-$$V$$-$$P$$-$$A$ in the bibliographic network, we construct a bipartite HIN consisting of $A$ and $V$, where the weight of edges between any pair of a type-$A$ vertex and a type-$V$ vertex equals to the numbers of path instances following $A$$-$$P$$-$$V$ between them.
    \end{itemize}

\HINSim refers to our proposed meta-path guided embedding model. Considering the choice of loss functions $\mathcal{L}_{\mathcal{M}, \mbox{seq}}$ and $\mathcal{L}_{\mathcal{M}, \mbox{pair}}$, the corresponding model are denoted as \HINSim-seq and \HINSim-pair respectively.

\noindent\textbf{Default Parameters.}
The parameter $\gamma$ controlling the effect of overly-popular vertices is set to $3/4$ inspired from~\cite{mikolov2013distributed}.
The dimension of the vertex embedding vectors, $d$, is set to $50$.
The negative sampling ratio $K$ is set to $5$, whereas the sampling times $t$ is set to $1$ million by default.
The number of working cores is set to $16$.
Talking about the initialization of global bias, local bias vectors, and embedding vectors, we assign all parameters as a random real value uniformly in $[-1, 1]$.
The learning rate in stochastic gradient descent is initialized as $0.25$ and later linearly decreased.

\noindent\textbf{Machine.} The following experiments on execution time were all conducted on a machine equipped two Intel(R) Xeon(R) CPU E5-2680 v2 @ 2.80GHz with 20 physical cores in total.
Our framework is fully implemented in C\texttt{++}\footnote{The source code will be published in the author's GitHub after acceptance.}.

\begin{table}[t]
        \centering
        \caption{AUC Evaluation on DBLP dataset.}
        \label{tbl:dblp}
        \scalebox{0.76}{
        \begin{tabular}{|c|c|c|c|}
        \hline
        Meta-path & Model & Research Area & Research Group \\
        \hline
        \hline
        \multirow{2}{*}{N/A} & \LINE-1st &52.32\%  & 52.89\%  \\
        & \LINE-2nd & 51.82\% &51.53\%   \\
        \hline
        \hline
        \multirow{4}{*}{$A$$-$$P$$-$$A$} & \PathSim &52.07\%  & \cellcolor{blue!25} 76.75\% \\
        & \PTE & 50.90\%  & \cellcolor{blue!25} 77.07\% \\
        & \HINSim-seq &52.97\%  &\cellcolor{blue!25} 73.97\% \\
        & \HINSim-pair &52.87\%  &\cellcolor{blue!25} \textbf{81.46\%} \\
        \hline
        \multirow{4}{*}{$A$$-$$P$$-$$V$$-$$P$$-$$A$} & \PathSim &\cellcolor{blue!25} 80.51\%  & 72.60\%\\
        & \PTE & \cellcolor{blue!25} 74.40\%  &65.87\%  \\
        & \HINSim-seq &\cellcolor{blue!25} 77.06\% & 74.83\% \\
        & \HINSim-pair &\cellcolor{blue!25} \textbf{83.58\%}  & 73.07\%\\
        \hline
        \multirow{4}{*}{$A$$-$$P$$-$$T$$-$$P$$-$$A$} & \PathSim & 55.22\%  & 61.16\% \\
        & \PTE & 68.38\% & 73.28\%  \\
        & \HINSim-seq &61.17\% &  65.73\% \\
        & \HINSim-pair &69.18\%  &74.96\%  \\
        \hline
        \hline
        Best Weighted & \HINSim-pair & \textbf{83.81\%} & \textbf{82.27\%} \\
        Combination              &              &0.1 {\small $A$$-$$P$$-$$A$} + & 0.9 {\small $A$$-$$P$$-$$A$} +  \\
        (grid search)             &              &0.9 {\small $A$$-$$P$$-$$V$$-$$P$$-$$A$} & 0.1 {\small $A$$-$$P$$-$$V$$-$$P$$-$$A$}\\
        \hline
        \end{tabular}
        }
    \vspace{-0.5cm}
    \end{table}

\subsection{AUC Evaluations}
\label{sec:auc}

Although it is hard to obtain the labels of the detailed rankings among all pairs of vertices, it is relatively easy to give an estimation based on the labels of the vertex groupings $l(\cdot)$.
Considering the ranking problem for each individual vertex $u$, if we rank the other vertices based on the similarity scores, it is very natural that we expect the vertices from the same group (similar ones) are at the top of the ranking list whereas the dissimilar ones are at the bottom of the list.
More specifically, we define the AUC score as follows. For a better similarity metric, the AUC score should be larger.
    \begin{eqnarray*}
    \scriptsize
    \begin{split}
        AUC = \dfrac{1}{|V|} \sum_{u \in V} \dfrac{\sum_{v, v' \in V \wedge l(u) = l(v) \wedge l(u) \neq l(v')} \mathbbm{1}_{sim(u, v) > sim(u, v')}}{\sum_{v, v' \in V \wedge l(u) = l(v) \wedge l(u) \neq l(v')} 1}
    \end{split}
    \end{eqnarray*}
Note that the models that generate similarity measures are learned from the whole dataset, whereas the AUC metric is calculated only in the subset of vertices where we have group labels. The subset is usually small because computing AUC needs pairwise similarities among the subset.

We have the following observations from Tables~\ref{tbl:dblp}~and~\ref{tbl:yelp}.
First, about single meta-path we have:
\begin{itemize}[noitemsep,nolistsep]
\item \emph{User-guidance is crucial.} The choice of the user-selected meta-path is really important and affects the performance of user-guided models significantly.
For example, $A$$-$$P$$-$$V$$-$$P$$-$$A$ works the best in the \textit{Research Area} grouping, where the shared venue is more telling. However, collaborations are more important in \textit{Research Groups} grouping, and thus $A$$-$$P$$-$$A$ works the best.
In the \textit{Yelp} dataset, although business names carry some semantics, words in review are more telling.
In both groupings, $B$$-$$R$$-$$W$$-$$R$$-$$B$ always better fits the user-preferred similarity.
These phenomena imply that a better fit meta-path will lead models such as \HINSim to better performance.
\item \emph{\HINSim-pair performs better than \HINSim-seq}, because the pairwise loss function exploits the meta-path guidance in a more thorough way. We will focus on \HINSim-pair in later experiments.
\item \emph{User-guided models perform better.} As long as the selected meta-path is reasonable (e.g., $A$$-$$P$$-$$V$$-$$P$$-$$A$ in the \textit{Research Area} grouping , $A$$-$$P$$-$$A$ in the \textit{Research Group} grouping, and $B$$-$$R$$-$$W$$-$$R$$-$$B$ in the \textit{Yelp} dataset), \HINSim, \PathSim, and \PTE always perform better than \LINE, which proves the importance of following the guidance from user-selected meta-paths.
\item \emph{\HINSim-pair performs the best} with significant advantages over \PathSim and \PTE, which demonstrates the power of embedding techniques and proper usage of network structures respectively.
\end{itemize}

\noindent\textbf{Meta-path Combination.} We choose the best performing meta-paths $A$$-$$P$$-$$A$ and $A$$-$$P$$-$$V$$-$$P$$-$$A$ in the \textit{DBLP} dataset and run grid searches for best weights to achieve highest AUC scores in \textit{Research Area} and \textit{Research Group} groupings respectively. Because the best performing meta-path in \textit{Yelp} dataset is always $B$$-$$R$$-$$W$$-$$R$$-$$B$, any combination is useless in this case.
Note that this grid search against grouping labels shows an upper bound of the highest possible AUC score, which can be rarely achieved without knowing labels.
As shown in Table~\ref{tbl:dblp}, the improvement of best weighted meta-paths is marginal. Therefore, weighted combination might be necessary to achieve the best performance but the choice of meta-paths is more important.

\begin{table}[t]
        \center
        \caption{AUC Evaluation on Yelp dataset.}
        \label{tbl:yelp}
        \scalebox{0.8}{
        \begin{tabular}{|c|c|c|c|}
        \hline
        Meta-path & Model & Business Type & Restaurant Type \\
        \hline
        \hline
        \multirow{2}{*}{N/A} & \LINE-1st & 78.89\%  &82.39\%  \\
        & \LINE-2nd &80.20\%  & 70.20\% \\
        \hline
        \hline
        \multirow{5}{*}{$B$$-$$R$$-$$W$$-$$R$$-$$B$} & \PathSim & \cellcolor{blue!25} 83.49\% & \cellcolor{blue!25} 74.66\%  \\
        & \PTE & \cellcolor{blue!25} 85.67\% &  \cellcolor{blue!25} 83.77\% \\
        & \HINSim-seq &\cellcolor{blue!25} 84.39\%& \cellcolor{blue!25} 78.62\%  \\
        & \HINSim-pair & \cellcolor{blue!25} \textbf{89.22\%} & \cellcolor{blue!25} \textbf{88.73\%}  \\
        \hline
        \multirow{5}{*}{$B$$-$$N$$-$$B$} & \PathSim & 53.77\% &55.26\%   \\
        & \PTE &  61.69\%  &  63.18\%\\
        & \HINSim-seq &62.53\% & 61.27\% \\
        & \HINSim-pair & 59.61\% & 59.39\% \\
        \hline
        \end{tabular}
        }
        \vspace{-0.5cm}
\end{table}


\begin{figure*}
	\centering
    \begin{minipage}{.6\textwidth}
      \centering
      \subfigure[\small Varying $d$.] {
        \includegraphics[width=0.3\textwidth]{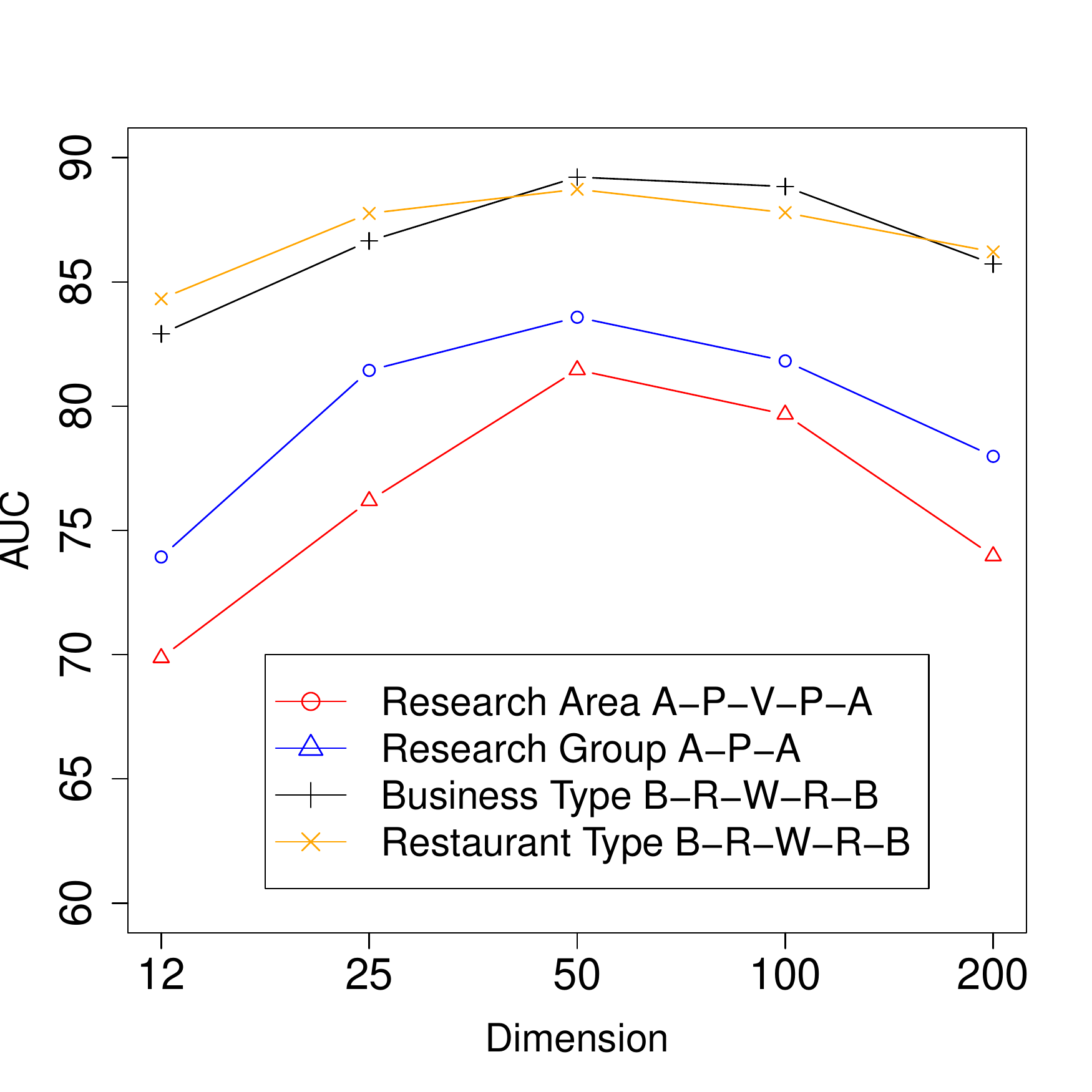}
        \label{fig:vary_d}
      }
      \subfigure[\small Varying $K$.] {
        \includegraphics[width=0.3\textwidth]{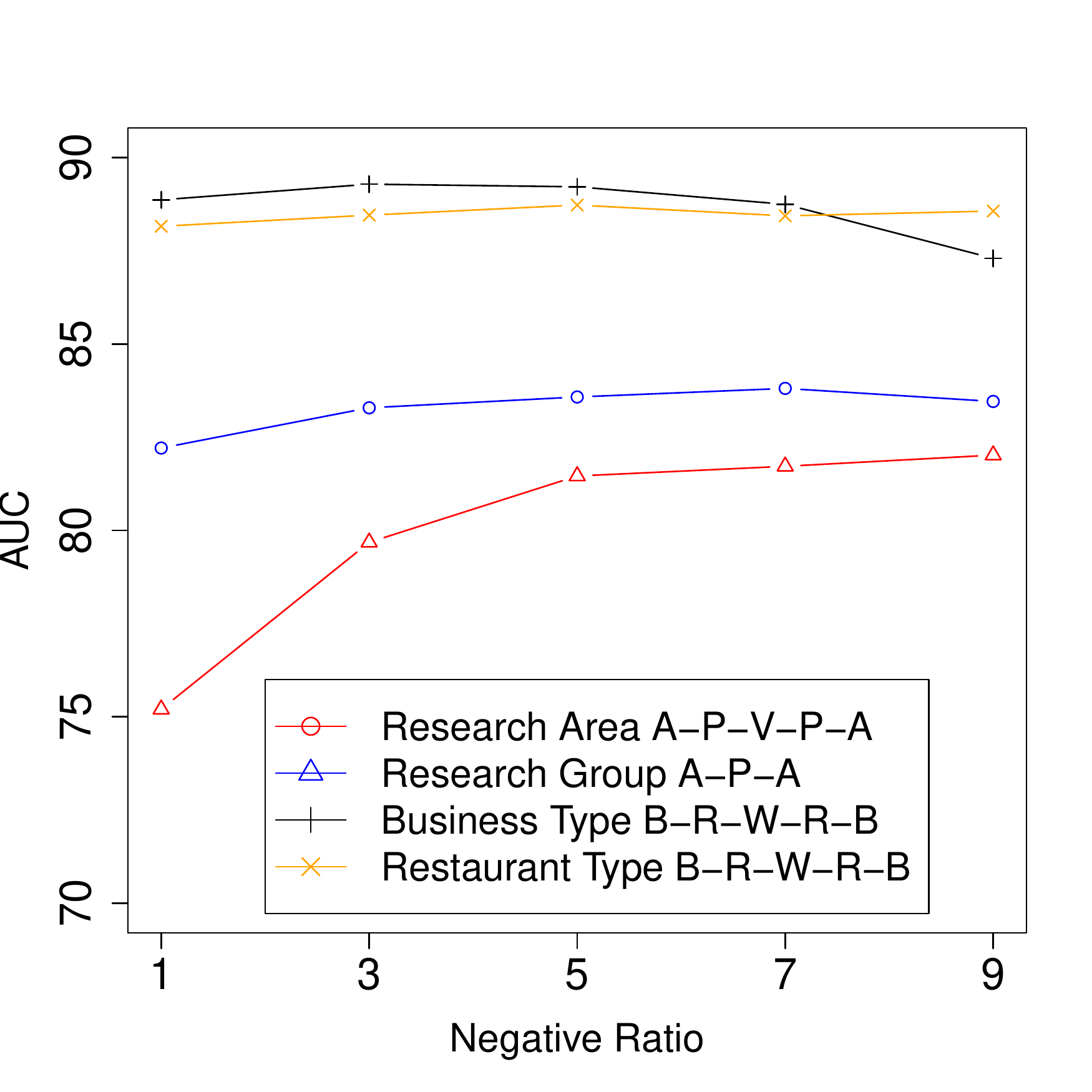}
        \label{fig:vary_k}
      }
      \subfigure[\small Varying total samples.] {
        \includegraphics[width=0.3\textwidth]{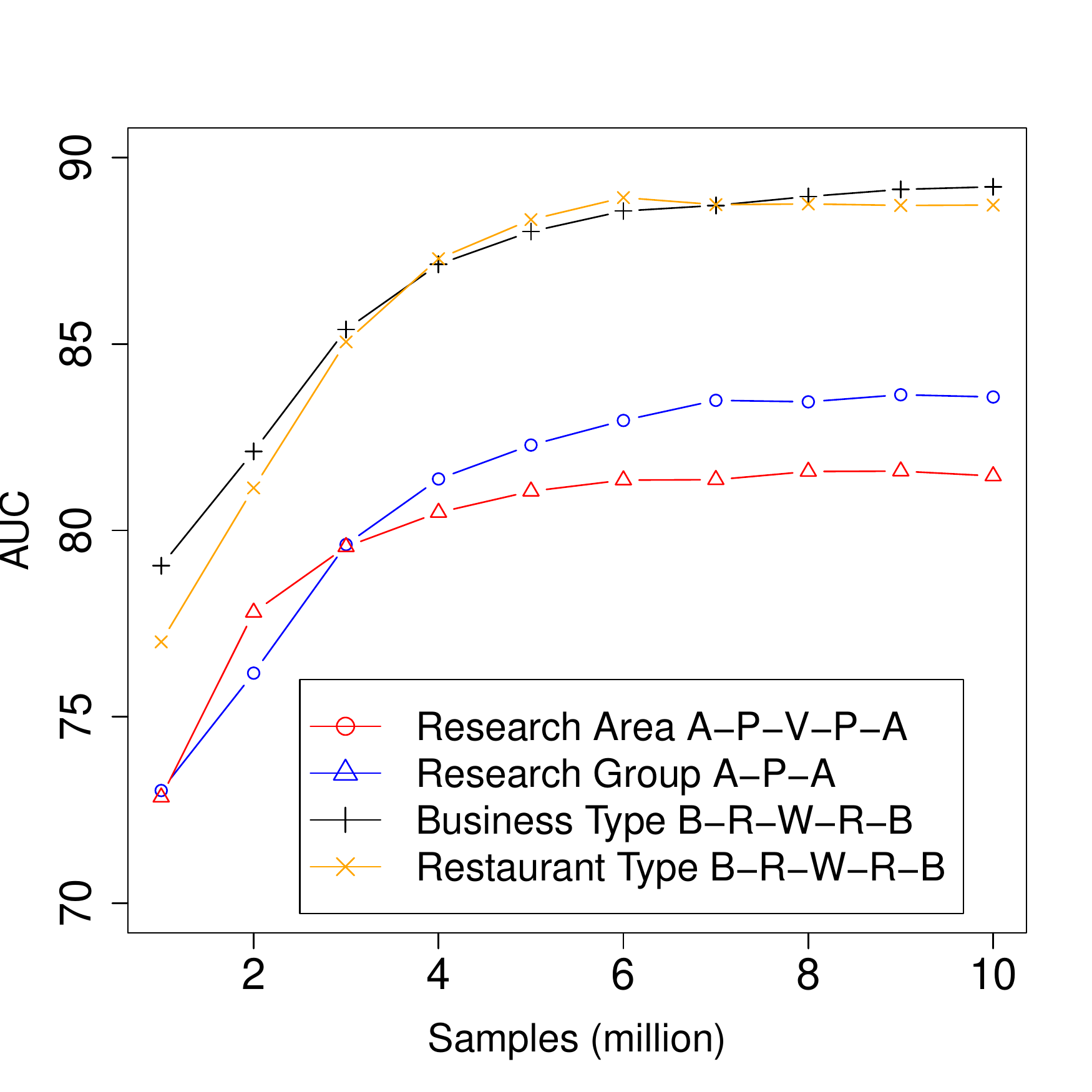}
        \label{fig:vary_sample}
      }
      \vspace{-0.4cm}
      \caption{Parameter Sensitivity of \HINSim-pair.}
	\end{minipage}%
	\begin{minipage}{.4\textwidth}
  	\centering
      \subfigure[\small Varying network sizes.] {
          \includegraphics[width=0.45\textwidth]{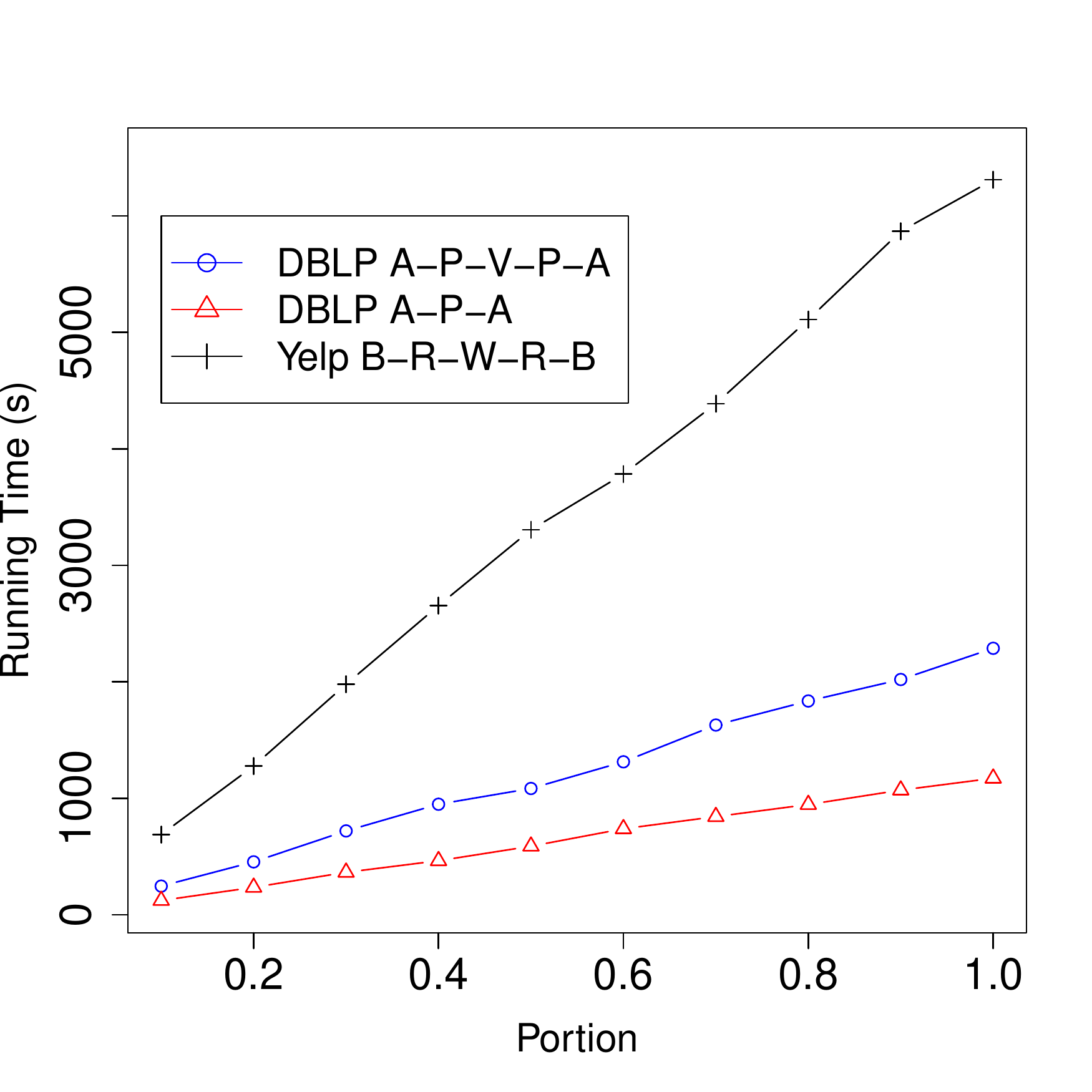}
          \label{fig:scale}
      }
      \subfigure[\small Varying cores.] {
          \includegraphics[width=0.45\textwidth]{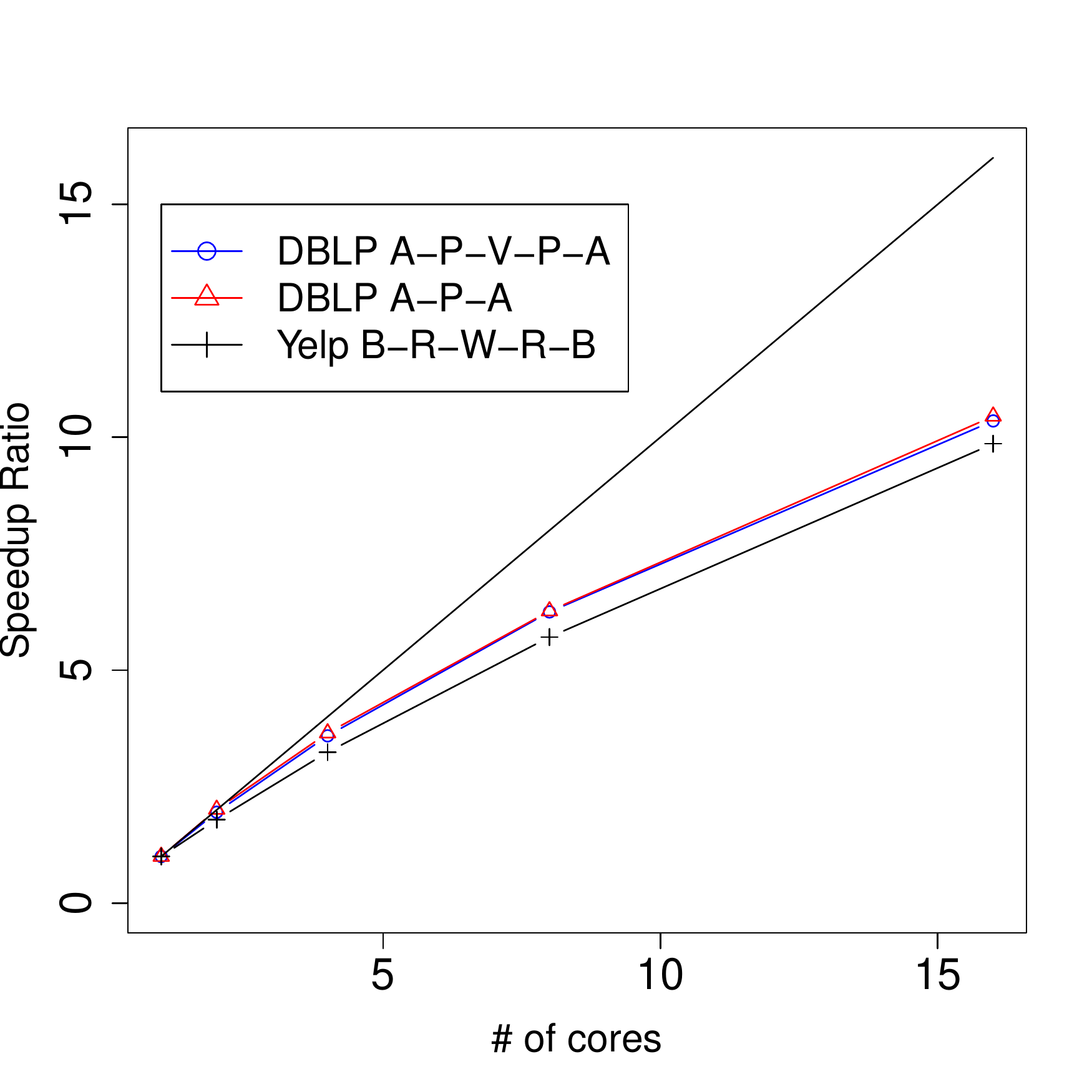}
          \label{fig:speedup}
      }
      \vspace{-0.4cm}
      \caption{Efficiency of \HINSim-pair.}
	\end{minipage}
	\vspace{-0.5cm}
\end{figure*}


\subsection{Visualizations}

With embedding vector for each vertex, we can show meaningful visualizations, which layout the vertices of the same type in a two-dimensional space, and check whether the boundaries between different groups are relatively clear.

Taking the \textit{DBLP} dataset as an example, we visualize the vertices with different colors regarding to their group labels in the \textit{Research Area} grouping and the \textit{Research Group} grouping.
Their embedding vectors are projected to a 2-D space using the t-SNE package~\cite{van2008visualizing}, which is a nonlinear dimensionality reduction technique and well suited for projecting high-dimensional data into a low dimensional space.

\begin{figure}[t]
      \centering
      \subfigure[\HINSim-pair] {
          \includegraphics[width=0.14\textwidth]{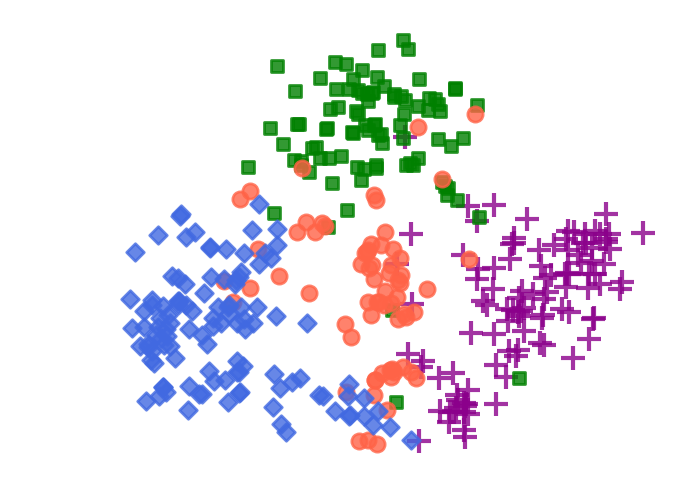}
      }
      \subfigure[\PTE] {
          \includegraphics[width=0.14\textwidth]{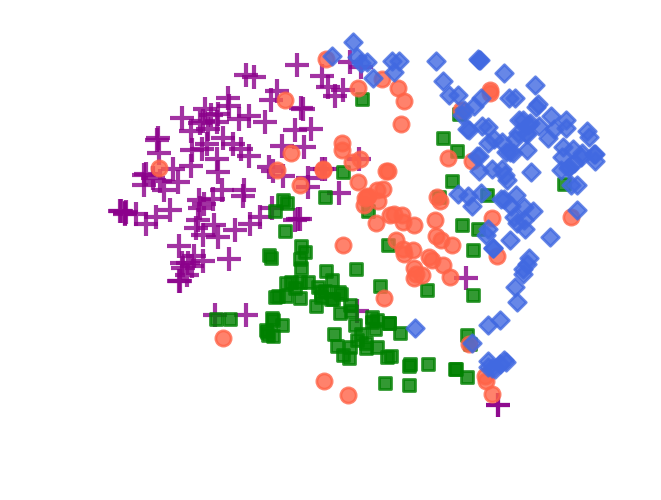}
      }
      \subfigure[\LINE-1st] {
          \includegraphics[width=0.14\textwidth]{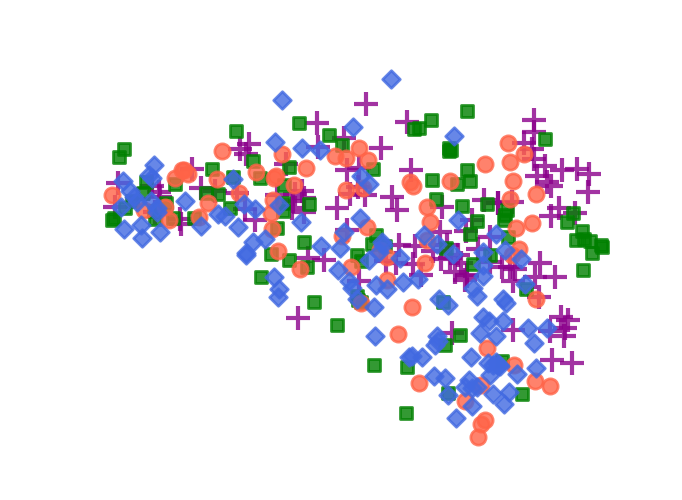}
      }
      \vspace{-0.4cm}
      \caption{Visualization of embedding vectors of $10\%$ random sampled authors in \textit{DBLP Research Area} grouping when $\mathcal{M}$$=$$A$$-$$P$$-$$V$$-$$P$$-$$A$. Colors correspond to research areas: \emph{\textcolor{blue}{Database}}, \emph{\textcolor{orange}{Data Mining}}, \emph{\textcolor{green}{Information Retrieval}}, and \emph{\textcolor{magenta}{Machine Learning}}.}
      \label{fig:vis_area}
      \vspace{-0.4cm}
\end{figure}

Laying out these vertex embedding vectors is challenging, especially for the vertices in the four closely related research areas: \mquote{data mining}, \mquote{database}, \mquote{machine learning} and \mquote{information retrieval}.
For different embedding-based methods (i.e., \HINSim, \PTE, and \LINE), we choose to visualize their variants holding the best AUC performance, i.e., \HINSim-pair using $A$$-$$P$$-$$V$$-$$P$$-$$A$, \PTE using $A$$-$$P$$-$$V$$-$$P$$-$$A$, and \LINE-1st.
As shown in Fig.~\ref{fig:vis_area}, we visualize $10\%$ random samples of $3,750$ authors from 4 research domains of computer science.
Better embedding vectors should lead to a clearer figure where the boundaries between different colored points should be clean and almost not interfering with each other.
Based on the visualizations, one can easily observe that \HINSim-pair using $A$$-$$P$$-$$V$$-$$P$$-$$A$ provides the best embedding vectors judged from this criterion.

\begin{figure}[t!]
  	\centering
      \subfigure[\HINSim-pair] {
          \includegraphics[width=0.14\textwidth]{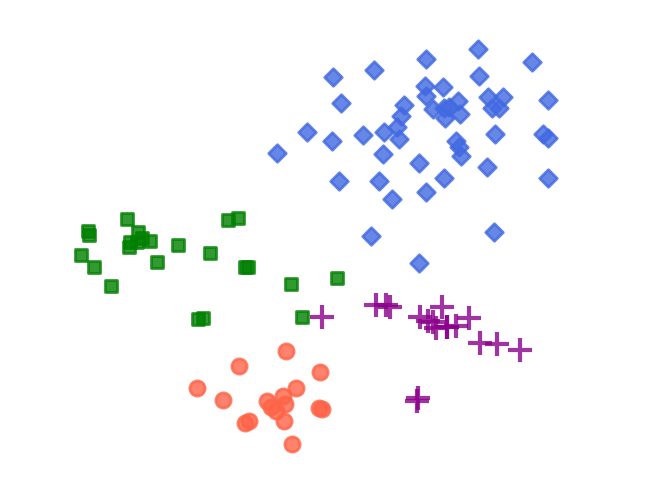}
      }
      \subfigure[\PTE] {
          \includegraphics[width=0.14\textwidth]{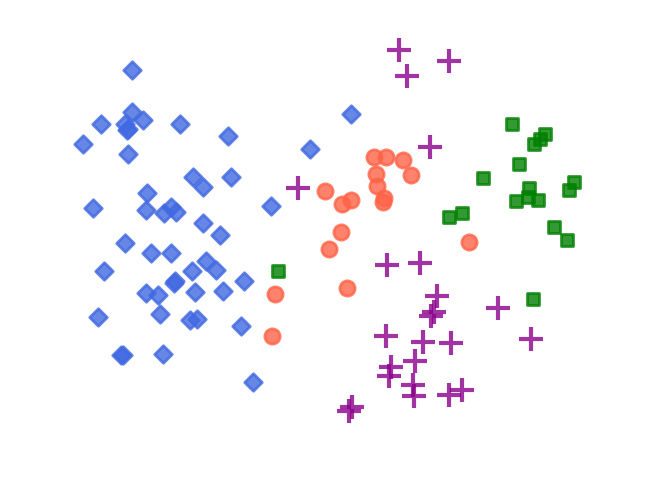}
      }
      \subfigure[\LINE-1st] {
          \includegraphics[width=0.14\textwidth]{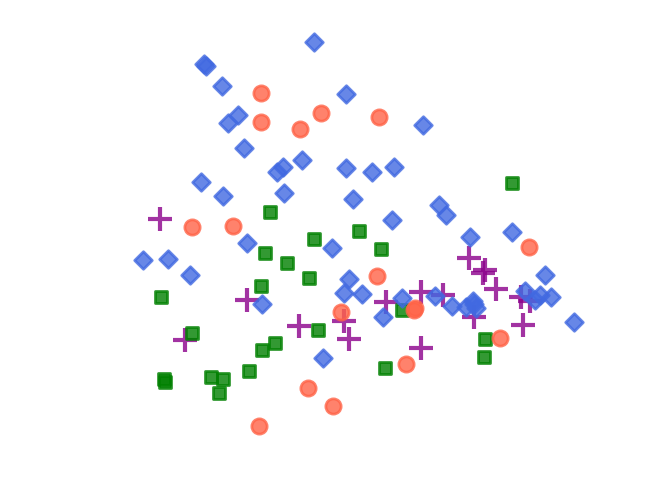}
      }
      \vspace{-0.4cm}
      \caption{Visualization of embedding vectors of all authors in \textit{DBLP Research Group} grouping when $\mathcal{M}$$=$$A$$-$$P$$-$$A$. Colors correspond to research groups: \emph{\textcolor{orange}{Christos Faloutsos}}, \emph{\textcolor{blue}{Jiawei Han}}, \emph{\textcolor{green}{Michael I. Jordan}}, and \emph{\textcolor{magenta}{Dan Roth}}.}
      \label{fig:vis_group}
      \vspace{-0.5cm}
\end{figure}

Similarly, the visualizations of the \textit{Research Groups} grouping based on the models with best AUC performance are shown in Fig.~\ref{fig:vis_group}. Our proposed model \HINSim-pair using $A$$-$$P$$-$$A$ clearly beats \PTE using $A$$-$$P$$-$$A$ and \LINE-1st.

The significant improvements over \PTE and \LINE observed via visualization are consistent with the previous evaluations.

\subsection{Parameter Sensitivity}
We select the best performing models \HINSim-pair to study the parameter sensitivity, such as using $A$$-$$P$$-$$V$$-$$P$$-$$A$ in the \textit{Research Area} grouping, using $A$$-$$P$$-$$A$ in the \textit{Research Group} grouping, and using $B$$-$$R$$-$$W$$-$$R$$-$$B$ in both \textit{Business Type} and \textit{Restaurant Type} groupings.
We vary the parameter values and see how the AUC performance changes.

Based on the curves in Fig.~\ref{fig:vary_d}, we can observe that setting the dimension ($d$) of vertex embedding vectors as $50$ is reasonable, because too small $d$ cannot sufficiently capture the semantics, while too large $d$ may lead to some overfitting problem.
Fig.~\ref{fig:vary_k} indicates that the AUC scores are not sensitive to the negative sample ratio $K$ and $K=5$ is a good choice.
As shown in Fig.~\ref{fig:vary_sample}, as more samples are optimized during training, the AUC scores keep an increasing trend and finally converge.

\subsection{Scalability and Parallelization}
We investigate the efficiency of \HINSim\ by considering both the scalability and the parallelization as shown in Fig.~\ref{fig:scale}.
We try different portions of network sizes (i.e., $|V| + |E|$) in the two networks and run our best performing models, i.e., \HINSim-pair using $A$$-$$P$$-$$V$$-$$P$$-$$A$ and using $A$$-$$P$$-$$A$ on the \textit{DBLP} dataset, as well as \HINSim-pair using $B$$-$$R$$-$$W$$-$$R$$-$$B$ on the \textit{Yelp} dataset.
Based on these curves, the running time is linear to the size of networks while the longer meta-path costs a little more time, which are consistent with our previous theoretical time complexity analysis.
We vary the number of working cores and run our models on the \textit{DBLP} and \textit{Yelp} datasets.
The results are plotted in Fig.~\ref{fig:speedup}.
The speedup is quite close to linear, which shows that \HINSim\ is quite scalable to the number of working cores.


\vspace{-0.3cm}

\section{Conclusions}\label{sec:con}

In this paper, we propose a general embedding-based similarity search framework for heterogeneous information networks (HINs).
Our proposed model, \HINSim, incorporates given meta-paths and network structures to learn vertex embedding vectors.
The similarity defined by the cosine similarity between vertex embeddings of the same type has demonstrated its effectiveness, outperforming the previous state-of-the-art algorithms on two real-world large-scale HINs.
The efficiency of \HINSim\ has also been evaluated and proved to be scalable.

There are several directions to further extend this work.
First, instead of similarities between vertices of the same type, one can also explore the relevances between vertices of different types.
Second, a mechanism could be developed to automatically learn and extract a set of interesting meta-paths or their weighted combinations from user-provided rankings or preferences.
Third, similarity is the fundamental operation for mining and exploring HINs.
This study on similarity measure, defined in HINs based on meta-path guided embedding, and its efficient computations will impact other searching and mining problems in HINs.
For example, it is necessary to re-examine clustering, classification and prediction functions in HINs by reconsidering the similarity measures defined based on meta-path guided embedding.
Also, mining outliers in networks can be formulated as finding a small subset of vertices with extremely low similarities to other vertices or clusters.

\newpage
\bibliographystyle{abbrv}
\bibliography{HINSim}

\end{document}